\documentclass[
 reprint,
superscriptaddress,
 preprintnumbers,
 nofootinbib,
 nobibnotes,
 bibnotes,
 amsmath,amssymb,
 aps,
 prl, 
 floatfix
]{revtex4-1}

\usepackage{graphicx}
\usepackage{dcolumn}
\usepackage{bm}
\usepackage{comment}
\usepackage{xcolor}
\usepackage{natbib}
\usepackage[nowatermark]{fixmetodonotes}
\usepackage{isotope}
\usepackage{verbatim}
\usepackage{siunitx}
\DeclareSIUnit[number-unit-product = ]\percent{\char`\%}

\usepackage{hyperref}
\usepackage[mathlines]{lineno}
\newcommand{\eV}{\mathrm{e\kern -0.1em V}}
\newcommand{\MeV}{\mathrm{Me\kern -0.1em V}}
\newcommand{\keV}{\mathrm{ke\kern -0.1em V}}

\begin{document}


\title{Measurement of the total neutron cross section on argon \\ in the 20 to 70 keV energy range} 

\author{S. Andringa}
\affiliation{Laborat{\'o}rio de Instrumenta{}{\c c}{\~a}o e F{\'i}sica Experimental de Part{\'i}culas (LIP), Av. Prof. Gama Pinto, 2, 1649-003, Lisboa, Portugal}

\author{Y.Bezawada}
\affiliation{University of California at Davis, Department of Physics and Astronomy, Davis, CA 95616, U.S.A.}

\author{T. Erjavec}
\affiliation{University of California at Davis, Department of Physics and Astronomy, Davis, CA 95616, U.S.A.}

\author{J. He}
\affiliation{University of California at Davis, Department of Physics and Astronomy, Davis, CA 95616, U.S.A.}

\author{J. Huang}
\affiliation{University of California at Davis, Department of Physics and Astronomy, Davis, CA 95616, U.S.A.}

\author{P. Koehler}
\affiliation{Los Alamos National Laboratory, Physics Division, Los Alamos, NM 87545, U.S.A.}

\author{M. Mocko}
\affiliation{Los Alamos National Laboratory, Physics Division, Los Alamos, NM 87545, U.S.A.}

\author{M. Mulhearn}
\affiliation{University of California at Davis, Department of Physics and Astronomy, Davis, CA 95616, U.S.A.}

\author{L. Pagani}
\affiliation{University of California at Davis, Department of Physics and Astronomy, Davis, CA 95616, U.S.A.}

\author{E. Pantic}
\affiliation{University of California at Davis, Department of Physics and Astronomy, Davis, CA 95616, U.S.A.}

\author{L. Pickard}
\affiliation{University of California at Davis, Department of Physics and Astronomy, Davis, CA 95616, U.S.A.}

\author{R. Svoboda}
\affiliation{University of California at Davis, Department of Physics and Astronomy, Davis, CA 95616, U.S.A.}

\author{J. Ullmann}
\affiliation{Los Alamos National Laboratory, Physics Division, Los Alamos, NM 87545, U.S.A.}

\author{J. Wang}
\affiliation{University of California at Davis, Department of Physics and Astronomy, Davis, CA 95616, U.S.A.}
\affiliation{South Dakota School of Mines and Technology, Physics Department, Rapid City, SD 57701 USA}

\collaboration{ARTIE Collaboration}

\date{\today}

\nocite{*}

\begin{abstract}

The cross section for neutron interactions on argon is an important design and operational parameter for a number of neutrino, dark matter, and neutrinoless double beta decay experiments which use liquid argon as a detection or shielding medium.  There is a discrepancy between the evaluated total cross section in the $20$ to $70~\rm$\,keV neutron kinetic energy region given in the ENDF database and a single measurement conducted by an experiment with a thin target (0.2 atoms/barn) optimized for higher cross sections. This gives rise to significant uncertainty in the interaction length of neutrons in liquid argon. This discrepancy is now resolved by new results presented here from the Argon Resonance Transport Interaction Experiment (ARTIE), a thick target experiment (3.3 atoms/barn) optimized for the small cross sections in this energy region.
\end{abstract}



\maketitle
\section{\label{sec:intro}Introduction}
Liquid argon (LAr) is used in a wide range of particle physics experiments
investigating neutrinos~\cite{Antonello:2013ypa, Acciarri:2016smi,
  Abi:2017aow, Acciarri:2016ooe}, dark matter~\cite{DEAP:2019,
  Aalseth:2017fik}, and neutrinoless double beta
decay~\cite{Ackermann:2012xja, Abgrall:2017syy}.  Achieving
the scientific objectives sought by these experiments relies on
understanding the transport of neutrons through LAr at a
level of precision which has only recently emerged as a critical
experimental requirement. \hfill \break
Recent studies have shown that understanding the behavior of neutrons presents a special challenge in liquid-argon-based experiments \cite{Ankowski_2015, Friedland_2019, Gardiner_2021}. 

\indent The neutron-argon total cross section from the ENDF~\cite{ENDF} evaluation has a destructive
interference feature at $57$\,keV, which appears as a dip in the cross section around this energy, where
interaction length in LAr, for a natural abundance of
isotopes, is $30$\,m.  This is important, as at this energy scale, neutrons only lose a small
fraction of their kinetic energy in single elastic collisions with 
relatively massive argon nuclei. Thus, even neutrons with kinetic
energy well above $57$\,keV have a significant probability of reaching the low cross section region
with the resulting long interaction length.  
The results of the
most recent previous measurement~\cite{Winters:1991}, contained in the EXFOR database\cite{exfor:2014}, are inconsistent
with ENDF evaluation in the region of this feature, with an
inferred interaction length of $4.2$\,m.  This discrepancy makes it
impossible to reliably predict the performance of LAr in transporting and/or shielding neutrons.
This paper presents the results of the
Argon Resonant Transport Interaction Experiment (ARTIE) which was designed specifically to resolve the neutron cross section discrepancy.

\section{\label{sec:setup}Experimental Method}

The transmission through a target in a neutron beam 
 $T(E)$ is defined as the fraction of neutrons in a medium which pass
through a distance $d$ without scattering.  This is related to the cross section by the equation:
\begin{equation}
  \sigma(E) = -\frac{m}{\rho_{eff}\,d}\,\ln \, T(E)
\label{eq:sigma}
\end{equation}
Where $m$ is the mass of an argon atom, $d$ is the target thickness, and $\rho_{eff}$ is the effective density. The ARTIE target was designed with a thickness
approximately twenty times larger than used in Ref.~\cite{Winters:1991}.  Thus, the inconsistent cross section values reported by ENDF and Ref. \cite{Winters:1991} result in a  20\% difference in $T(E)$ at the dip energy. 
While well-suited to the dip region, the target
becomes essentially opaque at energies where the cross section is higher, restricting our 
energy Region Of Interest (ROI) to $20$ to $70$\,keV. To make the measurement, Flight Path 13 (FP13) at the Lujan Neutron Scattering Center~\cite{LANSCEBeam:1990} was used. FP13 has a total flight path of about 64 meters, allowing for excellent Time Of Flight (TOF) energy resolution up to several hundred keV.

\begin{figure}[h]
    \centering
    \includegraphics[width=\linewidth]{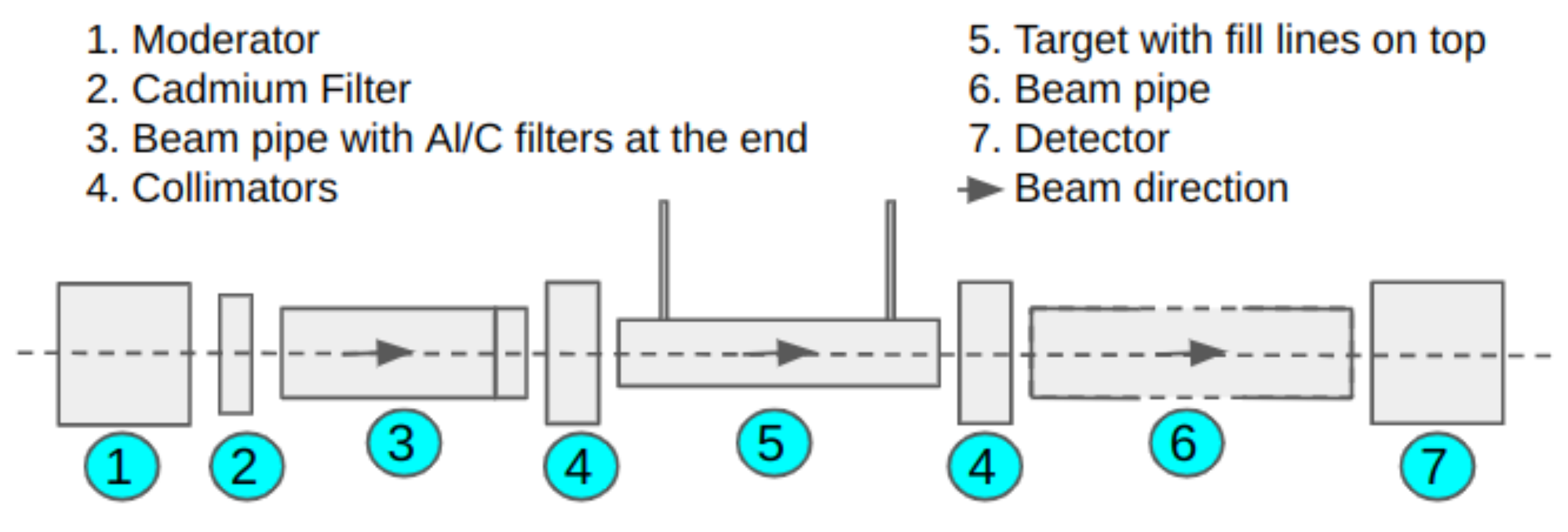}
    \caption{Configuration of the beam line elements. 1) Mark III LANSCE target and modrator. 2) cadmium filter for suppression of thermal neutrons. 3) $\sim 30\; m$ beam pipe where carbon and aluminum filters were mounted. 4) brass collimators. 5) ARTIE target. 6) $\sim 30\; m$ beam pipe. 7) 6-Li glass detector at $64\; m$ followed by beam dump.}
    \label{fig:config}
\end{figure}

The ARTIE experimental configuration is shown in Fig.~\ref{fig:config}. The ARTIE target consisted of a column of LAr of length $168$\,cm and diameter $25$\,mm, held at atmospheric pressure, and
contained in a vessel constructed from standard components.  
The target was inserted into FP13 at a distance of about $31$\,m from the upper-tier liquid hydrogen moderator of the
proton-accelerator-driven pulsed neutron source. 
The proton beam current (proportional to the resulting neutron flux) was monitored by a Current Transformer
(CT), with the integral output recorded every minute for relative beam normalization. 

The ARTIE neutron detector was located at the end of the beam line, about 
$30$\,m downstream of the target and consisted of a $9$\,cm
diameter by $1$\,mm thick $^6$Li-glass scintillator, viewed edge-on
by two RCA 8854 five-inch photomultiplier tubes (PMTs).  A triggered event was defined as a pulse above threshold in either PMT.  The
data acquisition system (DAQ) recorded the time and integrated PMT charge for each trigger. 

Two-inch thick brass cylinders with $6$\,mm holes in
the center were used to collimate the beam through the target.  Two
collimators were located upstream of the target, and two downstream.
These constrained the beam to the center of
the $25$\,mm diameter target, and produced a beam spot with an
$8$\,cm diameter at the $9$\,cm diameter neutron detector. Final alignment of the collimators was performed by first maximizing the DAQ-reported event rate. Collimators were then fine-tuned in order to produce a symmetric, fully contained image on storage-phosphor image plates at the detector location.

For insulation, the target was covered with a  rigid foam designed for cryogenic applications.  Near each end an upward opening feeds into commercial nalgene
dewars.  As argon boiled in the target, Gaseous Argon (GAr) vented through
the openings and was replaced by LAr from the dewars.  During operation, dewars were refilled about once per hour to ensure that the target remained full.  Data was collected during a two-week period, with most runs being in either {\it target-in} (LAr fill) or {\it target-out} (GAr fill) mode.  The use of GAr during target-out runs was accounted for by defining the effective target density as $\rho_{\rm eff}= \rho_{\rm in}-\rho_{\rm out}$. 
The transmission
was then experimentally determined as:
\begin{equation}
T(E) = \frac{N_{\rm in}(E) - B_{\rm in}}{N_{\rm out}(E) - B_{\rm out}}\cdot\frac{Q_{\rm out}}{Q_{\rm in}}
\label{eq:T}
\end{equation}
where $N_{\rm in/out}$ is the number of neutrons
observed during target-in/out runs, 
$Q_{\rm in/out}$ is the time-integrated beam current
 from the  CT monitor, and $B_{\rm in/out}$
is the experimentally determined background rate.

The collected data was subjected to
both run quality and individual event selection
cuts. Firstly, to minimize the
impact of potential non-linearity between the CT sensor and the neutron beam
intensity, the analysis only included the $95\%$ of collected data
taken while the CT was near the maximum value.
Secondly, data taken around the time of target filling were rejected since 
during these times, a roughly 30\% excursion in the beam-intensity normalized neutron event rate was
observed, which we surmise was caused by the unavoidable spilling of LAr vapor into the brass collimators from the filling dewar located above them. The rate
returned to the nominal value about 15 minutes after the end of the fill. Thus, these filling times were removed from the analysis by requiring that the post-fill rate return to at least 95\% of the pre-fill plateau value. This cut removed  about 12\% of the target-in data. Thirdly, both detector PMTs were required to have pulses within a 
$100$\,ns coincidence window, and to pass a cut which removed re-triggered events ({\it i.e.} two triggers from a single event). This last cut removed a negligible number of actual signal events. Following these cuts, there were 197k events recorded in the ROI for target-in runs, and 85k for target-out. 

\section{Calibration}
\label{sec:ecal}

{\it Energy:} The neutron detector recorded a time ($t_n$) and the proton beam pulse provided a start time ($t_0$). Together these gave a TOF which was used to determine the velocity ($v$) and hence the kinetic energy ($E$) of each event. These times were corrected for the average time a neutron scattered inside the moderator ($t_{\rm mod}$), usually referred to as the {\it Moderator
Function} (MF). The MF had been previously determined for LANSCE via
Monte Carlo simulation~\cite{moderator}. In the ROI this correction is typically 1-2\% with a smearing of roughly 10\% about the mean. 
The velocity $v$ is then a function of the TOF $t \equiv t_n-t_0$: 
\begin{equation}
  v(t) = \frac{L_{\rm fit}}{ t - t_{\rm mod}(t) + t_{\rm fit} } 
\label{eq:v(t)}
\end{equation}
where $L_{\rm fit}$ and $t_{\rm fit}$ were parameters determined by
fitting LAr data to known resonances of
aluminum and cadmium (both present in the beam line) and argon. The best fit value of $L_{\rm fit}$ was $63.82\pm 0.06$\,m which
agrees well with physical measurements made along the beam line.  The parameter $t_{\rm fit}$ (best fit $420\pm 29$\,ns) accounted for time delays in the detector, cables, and DAQ, as well as any residual
difference between the actual and simulated moderator response. In addition to the MF time-smearing,  the incident triangular-shaped neutron pulse had a
FWHM of $125$\,ns which led to a $53$\,ns uncertainty in $t_0$. These two factors dominated the energy resolution.

{\it Target Density:} Since the LAr in the target was always slightly boiling, there was always a small fraction of GAr present, which affected the overall target density. Thus, a separate experiment to directly measure the density of the liquid-gas mixture {\it in situ} was performed.
The mass of the target assembly ($M$) as a function of dewar liquid height ($h$) is given by:
    $M(h)=M_0+\left( \rho_{eff}-\rho_{air}  \right) V(h)
$   where $\rho_{eff}$ is the effective density of the argon mixture, $\rho_{air}$ is the density of air,  $M_0$ is the mass of the empty target, and $V(h)$ is the volume of the target as a function of liquid height. 
$M_0$ was measured using a precision scale ($\pm 1$\,g in the range of the roughly $25$\,kg target assembly mass), and $V(h)$ was determined by filling the target with known amounts of water and noting the level on a steel ruler inside the dewar. The dry, empty target was then filled with liquid argon while sitting on the precision scale.  During the subsequent boil off, a camera was used to simultaneously record the scale mass $M$ and the liquid level $h$, which were then analyzed to give $M(h)$. The observed boil off rate of $1.56$\,L/hr during this test was consistent with that observed during the actual neutron beam runs. A target-in density of $\rho_{in } =  1.318 \pm  0.017$\,g/cm$^3$ was obtained, which included a correction for ice-buildup on the target. This is $5.9$\% lower than the nominal density~\cite{NBS74} of pure liquid phase, and implies that this fraction of argon gas was mixed in the target during the beam runs. This resulted in an effective density $\rho_{eff}=3.30\pm 0.04$\, atoms/barn. Additionally, a 1.3\% upward adjustment in LAr density was made to account for the difference in density at the altitude of Los Alamos ($2300$\,m) as compared to the lab where the density test was performed ($16$\,m).  The uncertainty in the density measurement was taken into account when calculating the overall experimental uncertainty.


\section{\label{sec:analysis}BACKGROUNDS AND UNCERTAINTIES}

{\it Backgrounds:} Measurement of the transmission from
Eq.~\ref{eq:T} relied on subtracting background events, especially for the GAr runs. The two major backgrounds were: (i)  scattered neutrons from times earlier than the ROI that hit the detector at random later times, and (ii) gammas from neutron capture on water in the moderator. Other backgrounds such as ``wrap-around'' neutrons, prompt moderator gammas, and random backgrounds from other beam lines were determined to be insignificant.


To measure these major backgrounds, a standard technique \cite{Brown:2017accapp}\cite{MUCCIOLA2022100} that utilizes resonances that scatter or absorb nearly all incident neutrons was used.
Many beam line components are made of aluminum, and thus there are aluminum resonance features present in the data at $5.9$, $35$ and $88$\,keV. In addition, dedicated runs were made with an additional $2.54$\,cm aluminum filter inserted into the beam to further enhance these features. The background was then extracted at the resonance energy by subtracting the expected counts given the calculated aluminum transmission from the observed counts, given by

\begin{equation}  
  B_{out}(E)=\frac{N_{in,f}(E)-R_{in,f/out}(E)\,N_{out}(E)\,T(E)}{R_{in,f/out}(E)\left(1-T(E)\right)}
\label{eq:bkg}
\end{equation}
where $T(E)$ is the calculated effective transmission of neutrons with the aluminum filter, $N(E)$ are the counts in the bins at the resonance dips at $5.9$, $35$, and $88$\,keV for the filtered case ($in,f$) and unfiltered case ($out$), and $R_{in,f/out}(E)$ is the ratio of filtered background to the unfiltered background. This ratio is introduced to correct background reduction due to filter attenuation. 


Both the random-scatter neutron and the moderator capture gammas are seen to be nearly flat in time within our ROI \cite{Stamat:2022} for our neutron beam. As the filter attenuation effect is non-negligible for the $2.54$\,cm aluminum filter, it is corrected by $R_{in,f/out}(E)=B_{in,f}(E)/B_{out}(E)$ using an external measurement of the background components from \cite{Stamat:2022} and private communication with the authors \cite{privCommKoehler}. The ratio is $R_{in,f/out}(E_{ROI})=0.755_{-0.004}^{+0.021}$, with total uncertainties determined by applying Eq.~\ref{eq:bkg} for situations when $B_{out}(E)$ consists entirely of gammas or neutrons.

For LAr fill runs, gammas from the moderator were heavily suppressed due to target thickness. A small background still remained, and thus an argon resonance at $77$\,keV was used to evaluate this near the ROI. This background is also expected to be flat in TOF.

 \begin{figure}[tb]\centering
\includegraphics[width=\linewidth]{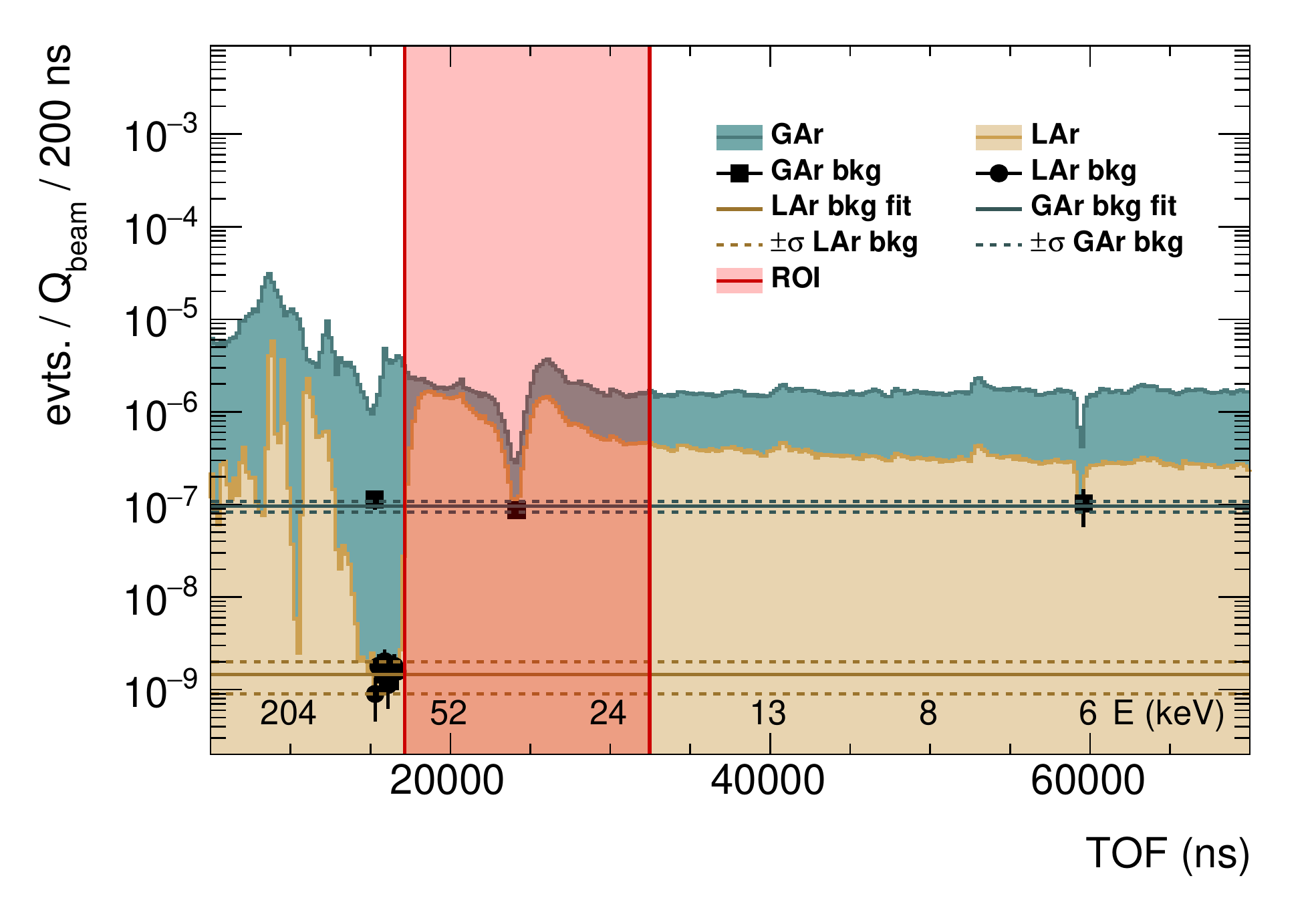}
\caption{The event rates and energy for a LAr and GAr filled target as a function of TOF. Filled points are the extracted background rates. 
}
\label{fig:bkg}
\end{figure}
Fig.~\ref{fig:bkg} shows the GAr and LAr event rates as a function of TOF and energy. The background rates are fit to a constant resulting in a background contribution relative to the signal in the ROI of about 0.14\% for LAr and 7.1\% for GAr, where the one-sigma uncertainties (dotted lines) are given by the flat fit to all background measurements. 


{\it Long-term Beam Line Stability:} 
The cross section calculation relies on the proportional nature between neutrons detected and the total integrated beam current, yet other than the $^6$Li detectors at the end of the beam line, there is no monitor of the neutron flux once the proton beam strikes the tungsten target. Since gaseous and LAr data runs involve different target setups, the uncertainties associated with the neutron beam in the target hall, after the neutrons are created, are not canceled naturally in Eq.\ref{eq:T}. To assess these uncertainties, event rates normalized by beam current were analyzed as a function of time for both air and LAr data, discussed below.

{\it Air:} The target was periodically run with air inside instead of liquid or gaseous argon. These air data, due to similar densities and neutron cross sections, was used as a surrogate for long-term beam line stability of GAr data. This assessment includes any systematic effects from changing atmospheric pressure. Three days of air data showed a daily modulation correlated closely to outside air temperature, and consistent across energies below, within, and above our ROI. Due to the close correlation to temperature, we suspect the effect is due to misalignment from the thermal expansion and contraction along the entire length of the beam line. The combined uncertainty on median air event rate is taken as an asymmetric systematic of $+3.14\%$ and $-3.93\%$ \cite{Erjavec2023}.

{\it LAr:} LAr data did not show the same modulation as the air data. A measurement of the event rate as a function of time for LAr includes event rate decrease from ice build-up on the kapton windows and also the cuts used to remove periods of refilling. The combined uncertainty on median LAr event rate is taken as an asymmetric systematic of $+0.69\%$, and $-1.06\%$ \cite{Erjavec2023}.



{\it Target Density:} As described in the calibration section, this was measured with a systematic uncertainty of
$1.3\%$.



After the selection cuts, several other systematic uncertainties were
determined to be negligible, including those from: non-linearity between
the beam intensity and CT measurement, dead time in the DAQ system \cite{Erjavec2023}, PMT afterpulsing \cite{Erjavec2023}, and contamination of the argon gas.

The absolute energy calibration was limited by the statistical
uncertainties on the fitted parameter $\delta(t_{\rm fit})=29~\rm$\,ns,
and the contribution from $\delta(L_{\rm fit})=0.064~\rm$\,m was
negligible.  No systematic uncertainty for energy resolution was applied
to the cross section results, which are reported here as a function of
measured energy.

The total systematic uncertainties on transmission and cross section are estimated for each energy point by using a toy Monte Carlo simulation where all relevant parameters are allowed to vary simultaneously around their central values. The resulting spreads  within the 68th percentile around the central-value measurements of transmission and cross section are taken as the total systematic uncertainties.

\section{\label{sec:conclusions}Conclusions}

\begin{figure}[tb]
\centering
    \includegraphics[width=\linewidth]{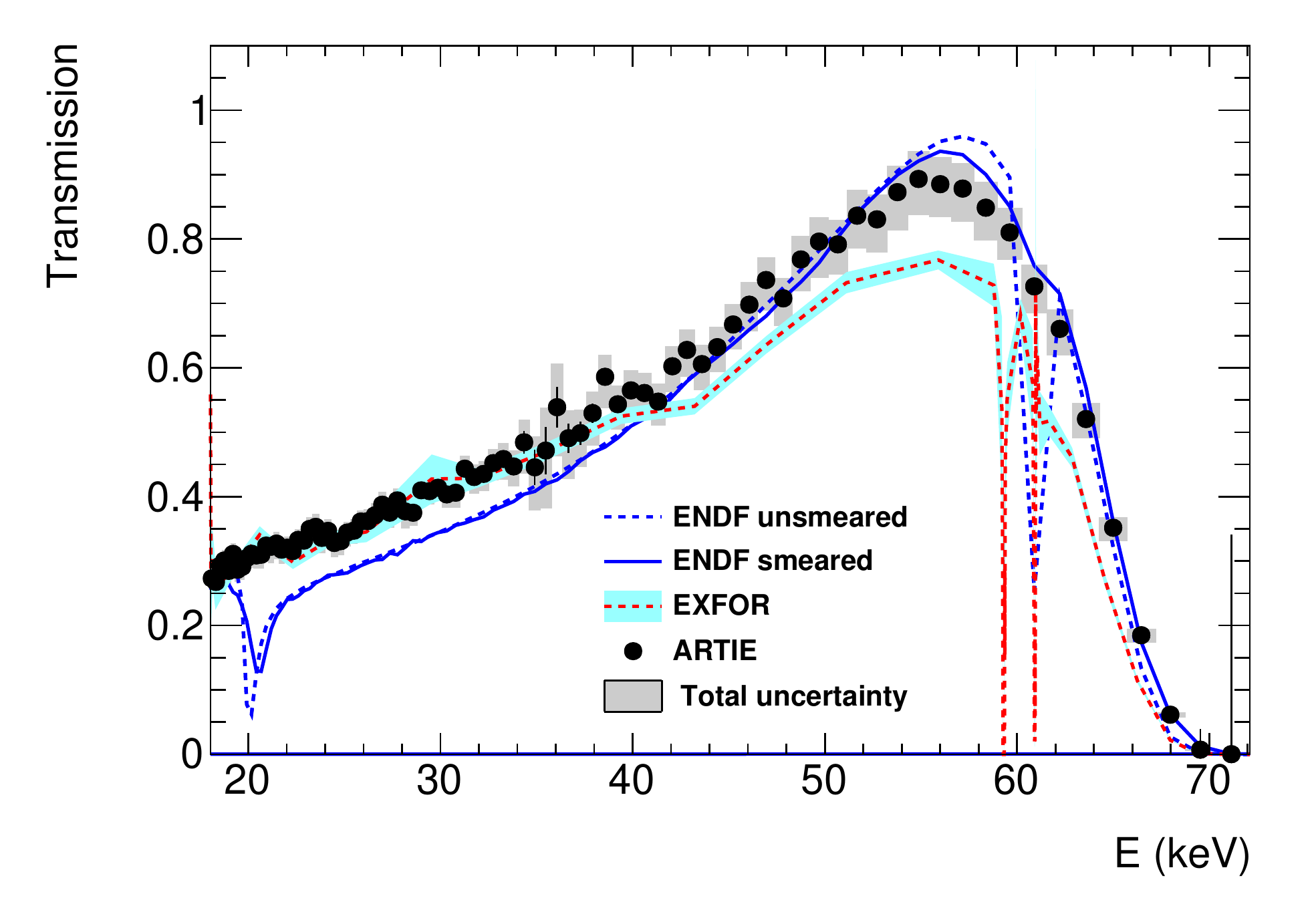}
\caption{The measured transmission of argon compared to EXFOR\cite{exfor:2014} and ENDF. The points are the central values and the bars represent the statistical uncertainty. The shaded grey regions are the total uncertainty, including systematics. For most energies, statistical error bars are smaller than the symbols.}
\label{fig:transmission}
\end{figure}

\begin{figure}[tb]
    \centering
    \includegraphics[width=\linewidth]{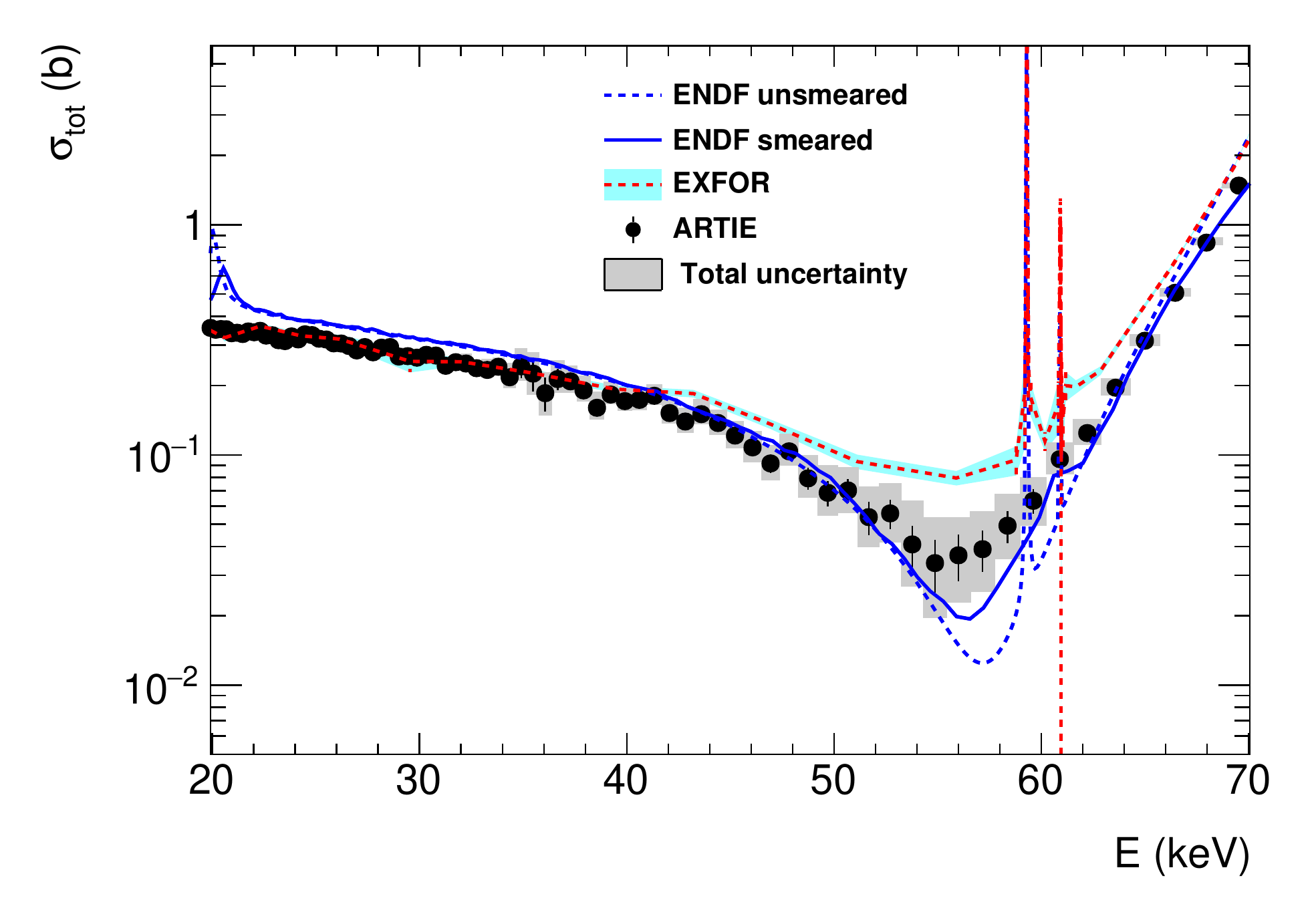}
    \caption{Neutron-argon total cross section as a function of energy. Also shown are the EXFOR evaluation and the ENDF evaluation smeared by the ARTIE energy resolution.}
    \label{fig:results}
\end{figure}

Using the TOF data and measured backgrounds, the
transmission $T$ is calculated from Eq.~\ref{eq:T} and the
cross section from Eq.~\ref{eq:sigma}.  The central value of each TOF
bin is converted to energy using Eq~\ref{eq:v(t)}. Fig.~\ref{fig:transmission} shows the transmission as a function of energy for the range $20$-$70$\,keV.
The measured neutron-argon total cross section as a function of
kinetic energy is shown in Fig.~\ref{fig:results}.
The dashed red and blue lines represent the EXFOR database \cite{exfor:2014} and ENDF evaluation, and the solid blue line is the ENDF prediction smeared by the ARTIE energy resolution. The beam energy resolution does not allow us to see the sharp features near $60$\,keV.
It can be seen that, for energies below 40\,keV the data is in good agreement with the EXFOR database,~\cite{Winters:1991}, while the ENDF evaluation shows slightly higher cross sections. Between 40 and 70\,keV, ARTIE data is in better agreement with ENDF, confirming the existence of the cross section dip. 
As a cross-check of our experimental setup and analysis technique,
the transmission for carbon was measured from data collected with
two $0.125\pm 0.010$” thick carbon (99.999\% purity)
disks~\cite{CarbonLesker} attached to the target while filled with GAr.  
There was good agreement between the measured ($0.73^{+0.03}_{-0.05}$) and predicted ($0.72$) transmission in the ROI with $\chi^{2}/\text{NDF} = 2.7 / 6 $, which confirms the analysis' methodology.\\
\indent In conclusion, our
results confirm a dip in the total cross section in the region of $50-60$\,keV. The point at $54.9$\,keV is found to have the lowest cross section of $\sigma=0.0339\pm {0.0087(stat.})_{-0.0112}^{+0.0175}(sys.) ~b$.
These results can now be used to reliably predict neutron transport to the level required in current and future experiments using liquid argon.

\section{\label{sec:acknowledgments}ACKNOWLEDGEMENTS}

Work at UC Davis was supported by the U.S. Department of Energy (DOE) Office
of Science under award number DE-SC0009999, and by the DOE National
Nuclear Security Administration through the Nuclear Science and
Security Consortium under award number DE-NA0003180. Support for LIP was from the Fundação para a
Ciência e a Tecnologia, I.P., project CERN/FIS-PAR/0012/2019. Major support was also
provided by the U.S. Department of Energy through the Los Alamos
National Laboratory. Los Alamos National Laboratory is operated by
Triad National Security, LLC, for the National Nuclear Security
Administration of U.S. Department of Energy (Contract
No. 89233218CNA000001).  Finally, we gratefully acknowledge the logistical and
technical support and the access to laboratory infrastructure provided
to us by LANSCE and its personnel.

\section{Bibliography}

\bibliographystyle{unsrt}
\bibliography{bibliography.bib}

\begin{thebibliography}{10}

\bibitem{Antonello:2013ypa}
M.~Antonello et~al.
\newblock {ICARUS at FNAL}.
\newblock 2013.

\bibitem{Acciarri:2016smi}
R.~Acciarri et~al.
\newblock {Design and Construction of the MicroBooNE Detector}.
\newblock {\em JINST}, 12(02):P02017, 2017.

\bibitem{Abi:2017aow}
B.~Abi et~al.
\newblock {The Single-Phase ProtoDUNE Technical Design Report}.
\newblock 2017.

\bibitem{Acciarri:2016ooe}
R.~Acciarri et~al.
\newblock {Long-Baseline Neutrino Facility (LBNF) and Deep Underground Neutrino
  Experiment (DUNE) Conceptual Design Report, Volume 4 The DUNE Detectors at
  LBNF}.
\newblock 2016.

\bibitem{Aalseth:2017fik}
C.~E. Aalseth et~al.
\newblock {DarkSide-20k: A 20 tonne two-phase LAr TPC for direct dark matter
  detection at LNGS}.
\newblock {\em Eur. Phys. J. Plus}, 133:131, 2018.

\bibitem{DEAP:2019}
R.~Ajaj et~al.
\newblock {Search for dark matter with a 231-day exposure of liquid argon using
  DEAP-3600 at SNOLAB}.
\newblock {\em Phys. Rev. D}, 100:022004, 2019.

\bibitem{Ackermann:2012xja}
K.~H. Ackermann et~al.
\newblock {The GERDA experiment for the search of $0\nu\beta\beta$ decay in
  $^{76}$Ge}.
\newblock {\em Eur. Phys. J.}, C73(3):2330, 2013.

\bibitem{Abgrall:2017syy}
N.~Abgrall et~al.
\newblock {The Large Enriched Germanium Experiment for Neutrinoless Double Beta
  Decay (LEGEND)}.
\newblock {\em AIP Conf. Proc.}, 1894(1):020027, 2017.

\bibitem{Friedland_2019}
Alexander Friedland and Shirley~Weishi Li.
\newblock Understanding the energy resolution of liquid argon neutrino
  detectors.
\newblock {\em Phys. Rev. D}, 99:036009, Feb 2019.

\bibitem{Ankowski_2015}
A.~M. Ankowski, P.~Coloma, P.~Huber, C.~Mariani, and E.~Vagnoni.
\newblock Missing energy and the measurement of the $cp$-violating phase in
  neutrino oscillations.
\newblock {\em Phys. Rev. D}, 92:091301, Nov 2015.

\bibitem{Gardiner_2021}
Steven Gardiner.
\newblock Simulating low-energy neutrino interactions with {MARLEY}.
\newblock {\em Computer Physics Communications}, 269:108123, dec 2021.

\bibitem{ENDF}
D.~A. Brown et~al.
\newblock {ENDF/B-VIII.0: The 8th Major Release of the Nuclear Reaction Data
  Library with CIELO-project Cross Sections, New Standards and Thermal
  Scattering Data}.
\newblock {\em Nucl. Data Sheets}, 148:1--142, 2018.

\bibitem{Winters:1991}
R.R. Winters, R.F. Carlton, C.H. Johnson, F.W. Hill, and M.R. Lacerna.
\newblock {Total cross section and neutron resonance spectroscopy for $n +
  ^{40}Ar$}.
\newblock {\em Phys. Rev. C}, 43:492, 1991.

\bibitem{exfor:2014}
V.Semkova N.Otuka, E.Dupont et~al.
\newblock {Towards a More Complete and Accurate Experimental Nuclear Reaction
  Data Library (EXFOR): International Collaboration Between Nuclear Reaction
  Data Centres (NRDC)}.
\newblock {\em Nuclear Data Sheets}, 120, 2014.

\bibitem{DUNETDR}
{B.~Abe and others}.
\newblock {Deep Underground Neutrino Experiment (DUNE), Far Detector Technical
  Design Report. FERMILAB-PUB-20-025-ND}.
\newblock 2020.

\bibitem{Abi2020DeepUN}
Babak Abi et~al.
\newblock Deep underground neutrino experiment (dune), far detector technical
  design report, volume iv far detector single-phase technology.
\newblock {\em arXiv: Instrumentation and Detectors}, 2020.

\bibitem{LANSCEBeam:1990}
P.W. Lisowski, C.D. Bowman, G.J. Russell, and S.A. Wender.
\newblock {The Los Alamos National Laboratory Spallation Neutron Sources}.
\newblock {\em Nucl. Inst. and Meth.}, 106:208, 1990.

\bibitem{moderator}
Lukas Zavorka, Michael~J. Mocko, Paul~E. Koehler, and John~L. Ullmann.
\newblock {Benchmarking of the MCNPX Predictions of the Neutron Time-emission
  Spactra at LANSCE}.
\newblock American Nuclear Society, 20th Topical Meeting of the Radiation
  Protection and Shielding Division of ANS, 2018.

\bibitem{NBS74}
H.M. Roder.
\newblock Liquid densities of oxygen, nitrogen, argon, and parahydrogen.
\newblock {\em NBS Technical Note 361 (Revised)}, 1974.

\bibitem{Brown:2017accapp}
J.M. Brown, A.~Youmans, N.~Thompson, Y.~Danon, D.P. Barry, G.~Leinweber, M.J.
  Rapp, R.C. Block, and R.~Bahran.
\newblock {Neutron Transmission Measurements and Resonance Analysis of
  Molybdenum-96}.
\newblock {\em AccApp '17}, 2017.

\bibitem{CarbonLesker}
Carbon (graphite) (c) sputtering targets.
\newblock
  \url{https://www.lesker.com/newweb/deposition_materials/depositionmaterials_sputtertargets_1.cfm?pgid=car1}.

\bibitem{Erjavec2023}
T.~Erjavec.
\newblock {\em {Understanding Neutron Transport for Liquid Argon Rare-Event
  Searches}}.
\newblock PhD thesis, University of California Davis, Davis, CA, 2024.

\bibitem{Stamat:2022}
A.~Stamatopolous, P.~Koehler, A.~Couture, B.~DiGiovine, G.~Rusev, and
  J.~Ullmann.
\newblock {New capability for neutron transmission measurements at LANSCE: The
  DICER instrument}.
\newblock {\em NIM A '22}, 2022.

\bibitem{privCommKoehler}
P.~Koehler.
\newblock {Private Communication}, 2022.

\bibitem{MUCCIOLA2022100}
R.~Mucciola, C.~Paradela, G.~Alaerts, S.~Kopecky, C.~Massimi, A.~Moens,
  P.~Schillebeeckx, and R.~Wynants.
\newblock Evaluation of resonance parameters for neutron interactions with
  molybdenum.
\newblock {\em Nuclear Instruments and Methods in Physics Research Section B:
  Beam Interactions with Materials and Atoms}, 531:100--108, 2022.

\end{thebibliography}

\end{document}